\documentclass[prd,superscriptaddress,amsfonts,amssymb,amsmath,showpacs,twocolumn]{revtex4}
\usepackage{epsfig}
\usepackage{graphics}  
\usepackage{color}
\usepackage{graphicx}
\usepackage[font={footnotesize,it}]{caption}
\usepackage[colorlinks=false,
            linkcolor=red,
            urlcolor=red,
            citecolor=blue]{hyperref}

\begin{document}

\title{Light Deflection by a Rotating Global Monopole Spacetime}

\author{Kimet Jusufi}
\email{kimet.jusufi@unite.edu.mk}
\affiliation{Physics Department, State University of Tetovo, Ilinden Street nn, 1200, Tetovo,
Macedonia}

\author{Marcus C. Werner}
\email{werner@yukawa.kyoto-u.ac.jp}
\affiliation{Yukawa Institute for Theoretical Physics, Kyoto University,
Kyoto 606-8502, Japan}

\author{Ayan Banerjee}
\email{ayan\_7575@yahoo.co.in}
\affiliation{Department of
Mathematics, Jadavpur University, Kolkata 700 032, West Bengal,
India}

\author{Ali \"{O}vg\"{u}n}
\email{ali.ovgun@emu.edu.tr}
\affiliation{Physics Department, Eastern Mediterranean University, Famagusta 99628, Northern Cyprus, 
Turkey}
\affiliation{Instituto de F\'{\i}sica, Pontificia Universidad Cat\'olica de
Valpara\'{\i}so, Casilla 4950, Valpara\'{\i}so, Chile}

\date{\today }

\begin{abstract}
We investigate the deflection of light by a rotating global monopole spacetime and a rotating Letelier spacetime in the weak deflection approximation. To this end, we apply the Gauss-Bonnet theorem to the corresponding osculating optical geometries and show that the deflection of light increases due to the presence of the global monopole parameter and the string cloud parameter, respectively. The results obtained for the deflection angle in the equatorial plane generalize known results for the corresponding non-rotating global monopole and Letelier spacetimes as well as the Kerr solution.
\end{abstract}


\maketitle

\section{Introduction}

Gravitational lensing is the bending of light trajectories between a background light source and us, the observer, due to massive objects (cf. \cite{Schneider} for a general treatment). It was first observed by Eddington's 1919 eclipse expeditions (cf. \cite{Eddington}), which provided an important test of general relativity in the solar system, and has now become one of the most useful tools in modern astronomy and cosmology for probing spacetime and the search for dark matter. Furthermore, it is hoped that gravitational lensing may eventually be used to test the fundamental theory of gravity and its possible modifications on cosmological scales, and shed light on the possible existence and properties of exotic compact objects.

One such application concerns topological defects. During phase transitions in the early universe, different types of topological objects may have formed, such as domain walls, cosmic strings and monopoles (e.g. \cite{Kibble}). A global monopole is a spherically symmetric gravitational topological defect with divergent mass which is thought to have arisen in the phase transition of a system composed of a self-coupling triplet of scalar fields $\phi^a$  which undergoes a spontaneous breaking of global $O(3)$ gauge symmetry down to $U(1)$. Gravitating global monopoles are stable against spherical as well as polar perturbations \cite{Watabe}.

The static solution of a global monopole was introduced in a classic paper by Barriola and Vilenkin \cite{birrola}. According to this model, global monopoles are configurations whose energy density decreases with the distance as $r^{-2}$ and whose spacetimes exhibit a solid angle deficit given by $\delta=8\pi^2 \eta^2$, where $\eta$ is the scale of gauge-symmetry breaking. Self-gravitating magnetic monopoles have also been investigated using numerical analysis \cite{Lee}. More recently, global monopoles have been discussed in spacetimes with a cosmological constant, e.g. in \cite{Bertrand}. Static spherically symmetric composite global-local monopoles have also been studied \cite{Achucarro}.

By applying a complex coordinate transformation, Newman and Janis \cite{Newman} established a relationship between the non-rotating and rotating spacetimes of general relativity. Using the method of a complex coordinate transformation, the rotating global monopole solution was obtained from the non-rotating counterpart solution by Teixeira Filho and Bezerra \cite{Bezerra1}. Adapting the same procedure, Morais Gra\c{c}a and Bezerra \cite{Graca} obtained the solution corresponding to a rotating global monopole from its static counterpart in the framework of $f(R)$ gravity theory.

Gravitational lensing in spacetimes with a non-rotating global monopole has been variously considered, for instance by Cheng and Man \cite{Cheng} who studied lensing in the strong field of a Schwarzschild black hole with a solid deficit angle owing to a global monopole. Recently, it has also been proposed that global monopole lensing effects may even be used to test Verlinde's emergent gravity theory \cite{liu}.

In this paper, we shall address the seemingly still open problem of calculating the deflection angle in a rotating global monopole-type spacetime.

Given a background spacetime, the propagation of light in can be studied by means of various techniques. Here, we propose to apply a geometrical method introduced by Gibbons and Werner \cite{gibbons1} which uses the optical geometry whose geodesics are the spatial light rays. By considering the Riemannian optical geometry of static spacetimes, it was shown how the asymptotic deflection angle can be computed using the Gauss-Bonnet theorem. This method was later extended by Werner \cite{werner} to stationary spacetime metrics (e.g. the Kerr black hole) whose optical geometry is Finslerian. Recently, it was investigated how this method may be used to calculate the deflection angle for finite distances in Schwarzschild-de Sitter and Weyl conformal gravity \cite{asahi1}; in the strong deflection limit \cite{asahi2}; and for charged black holes with topological defects and the spinning cosmic string in \cite{kimet}.

This method is interesting primarily from a conceptual point of view. Remarkably, the Gauss-Bonnet theorem allows the computation of the deflection angle by considering a domain \textit{outside} of the light ray, whereas the deflection is usually thought of as an effect due to mass primarily inside of the impact parameter. This underscores the global, partially topological nature of gravitational lensing. Thus, the method is well suited to the global monopole spacetime considered here.

Using the Gauss-Bonnet method, then, we shall compute the deflection of light in a rotating global monopole spacetime and a rotating Letelier spacetime in the weak deflection approximation. The paper is organized as follows: in Section II, we give a brief review concerning the the global monopole and rotating global monopole. Then we consider the Finsler optical geometry of the rotating global monopole spacetime in Section III, and proceed with the application of the Gauss-Bonnet method to its osculating Riemannian optical geometry in order to compute the deflection of light in section IV. We shall also discuss the Letelier spacetime, which represents a static and spherically symmetric black hole that is surrounded by a radially directed cloud of strings in Section V, followed by an analogous computation of the deflection of light in Section VI. A summary and discussion of our results is given in Section VII. Finally, in the Appendix A, we compare our method to a standard geodesic computation. In this paper, we shall use the natural units $c= G= \hbar = 1$ and the metric signature $(-, +, +, +)$. Greek indices refer to spacetime coordinates, and Latin indices denote spatial coordinates or those in optical geometry.

\section{Rotating Global Monopole Spacetime}
 A global monopole is a heavy object formed in the phase transition of a system composed by a self-coupling scalar triplet $\phi^{a}$. The simplest model which gives rise to a global monopole is described by the Lagrangian density \cite{Bezerra de Mello}

\begin{equation}\label{1-2}
\mathcal{L}=-\frac{1}{2}\sum_a g^{\mu\nu}\partial_{\mu}\phi^{a} \partial_{\nu}\phi^{a}-\frac{\lambda}{4}\left(\phi^{2}-\eta^{2}\right)^{2},
\end{equation}
with $a=1, 2, 3$, while $\lambda$ is the self-interaction term and $\eta$ is the scale of a gauge-symmetry breaking. The field configuration describing a monopole is
\begin{equation}
\phi^{a}=\frac{\eta f(r) x^{a}}{r},
\end{equation}
in which
\begin{equation}
x^{a}=\left\lbrace r \sin\theta \, \cos\varphi, r \sin\theta \,\sin\varphi,r \cos\theta \,\right\rbrace,
\end{equation}
such that $\sum_a x^{a}x^{a}=r^{2}$. It is interesting to note that, outside the core $f(r)\approx 1$, the energy-momentum
tensor is not zero and can be approximated as $T^{t}_{t}=T^{r}_{r}\simeq \eta^{2}/r^2$ and $T^{\theta}_{\theta}=T^{\varphi}_{\varphi}=0$. In a seminal paper \cite{birrola}, Barriola and Vilenkin have shown that the gravitational field of a global monopole black hole is described by the following spherically symmetric metric (see also \cite{vilenkin}, p. 424)
\begin{eqnarray}\label{4-2}\notag
\mathrm{d}s^2&=&-\left(1-\frac{2M}{r}\right)\mathrm{d}t^2+\frac{\mathrm{d}r^2}{\left(1-\frac{2M}{r}\right)}+\beta^2 r^2 \\
& \times & \left(\mathrm{d}\theta^2+\sin^{2}\theta \mathrm{d}\varphi^2\right),
\end{eqnarray}
where the global monopole parameter is given by $\beta^2=1-8\pi \eta^2$ and $M\approx M_{core}$ denotes the global monopole core mass, where $M_{core}\approx \lambda^{-1/2} \eta$ \cite{birrola}. For a typical grand unification scale $\eta=10^{16}$ GeV, which leads to $8 \pi \eta^2 \approx 10^{-5}$, while $\beta$ belongs to the interval $0<\beta\leq 1$. Recently, Filho and Bezerra, by applying the method of complex coordinate transformation, extended the static global monopole solution to a rotating global monopole spacetime metric. In particular, they found a metric with the following metric tensor components \cite{Bezerra1}
\begin{widetext}
\begin{equation}\label{5-2}
g_{\mu\nu}=\begin{bmatrix}
\vspace{0.3cm}
-\left(1-\frac{2Mr}{r^2+a^2 \cos^{2}\theta}\right) & 0 & 0 & -\frac{2 M a r\sin^{2}\theta}{r^2+a^2 \cos^{2}\theta} \\
\vspace{0.3cm}
0 & g_{rr} & 0 & g_{r\varphi}\\
\vspace{0.3cm}
0 & 0 & \beta^{2}(r^2+a^2 \cos^{2}\theta) & 0 \\
\vspace{0.3cm}
-\frac{2 M a r\sin^{2}\theta}{r^2+a^2 \cos^{2}\theta} & g_{\varphi r} & 0 & g_{\varphi\varphi}
\end{bmatrix},
\end{equation}
where
\begin{eqnarray}\notag
g_{rr}&=&\frac{r^2-a^2\left[\left(1-\beta^2\right)\sin^{2}\theta-\cos^{2}\theta\right]}{r^2-2Mr+a^2}-\left(1-\beta^2\right)\frac{a^2 \sin^{2}\theta \left[2Mr-a^2\left(1-\sin^{4}\theta\right)\right]}{\left(r^2-2Mr+a^2\right)^2},\\\notag
g_{r\varphi}&=&\left(1-\beta^2\right)\frac{a\left[r^2 \sin^{2}\theta-a^2 \cos^{2}\theta \left(1+\cos^{2}\theta\right)\right]}{r^2-2Mr+a^2},\\\notag
g_{\varphi\varphi}&=&\sin^{2}\theta \frac{\left\lbrace \beta^2 r^4 +\left[1-\left(1-2\beta^2\right)\cos^{2}\theta \right]a^2 r^2+2Ma^2 r \sin^{2}\theta +a^4 \cos^{2}\theta \left(\beta^2 \cos^2\theta+\sin^2 \theta \right)\right\rbrace} {r^2+a^2\cos^2 \theta}.
\end{eqnarray}

One can easily convince oneself that the Kerr solution in Boyer-Lindquist coordinates is recovered by setting $\beta=1$. If we consider only the linear terms in $a$, then we are left with the following stationary metric \cite{Bezerra1}
\begin{equation}\label{6-2}
\mathrm{d}s^2=-\left(1-\frac{2M}{r}\right)\mathrm{d}t^2+\frac{\mathrm{d}r^2}{\left(1-\frac{2M}{r}\right)}+\beta^2 r^2 \left(\mathrm{d}\theta^2+ \sin^{2}\theta \mathrm{d}\varphi^2\right)-\frac{4 M a \sin^{2}\theta}{r} \mathrm{d}t \mathrm{d}\varphi+\frac{2\,a\,(1-\beta^2)\sin^{2}\theta }{\left(1-\frac{2M}{r}\right)}\mathrm{d}r \mathrm{d}\varphi.
\end{equation}
\end{widetext}
Moreover, if we set $a=0$ in the last metric we recover the famous Barriola--Vilenkin static solution \eqref{4-2}. In the following, we are going to use the stationary metric \eqref{5-2} to calculate the deflection of light in the weak deflection approximation by applying the Gauss-Bonnet theorem to the corresponding osculating optical metric.

\section{Global monopole optical metric}

The stationary spacetime metric \eqref{5-2} gives rise to a Finslerian optical geometry of Randers type. Generally speaking, a Finsler metric $F$ on a smooth manifold $\mathcal{M}$ with $x\in \mathcal{M},\ X\in T_x M$ may be defined as a real, non-negative and smooth function of the tangent bundle away from the zero section (i.e. $X \neq 0$) which is positively homogeneous of degree one in the vectors $X$ and has a positive definite Hessian
\begin{equation}
g_{ij}(x,X)=\frac{1}{2}\frac{\partial^{2}F^{2}(x,X)}{\partial X^{i}\partial X^{j}}.\label{10-3}
\end{equation}
Then by homogeneity, $F^ {2} (x, X) =g_ {ij} (x, X) X^ {i} X^ {j} $, analogous to the square of the vector length in Riemannian geometry. Let us also briefly mention here that a Randers metric is a special Finsler metric that can be written as follows,
\begin{equation}
F(x, X)=\sqrt{a_{ij}(x)X^{i}X^{j}}+b_{i}(x)X^{i},\label{11-3}
\end{equation}
where $a_{ij}$ denotes a Riemannian metric and $b_{i}$ is a one-form satisfying the condition $a^{ij}b_{i}b_{j}<1$.

Now in order to see how the Randers optical metric arises from the rotating global monopole spacetime, let us recall that the general Randers form of a stationary spacetime can be written as \cite{gibbons2}
\begin{equation}\label{13-3}
\mathrm{d}s^2=V^2\left[-\left(\mathrm{d}t-b_i \mathrm{d}x^i \right)^2+a_{ij}\mathrm{d}x^i \mathrm{d}x^j\right].
\end{equation}
After some simple manipulations, we can recast the stationary global monopole metric \eqref{5-2} in the form of the Randers metric \eqref{13-3}. In particular, we find the following relations for $a_{ij}$ and $b_{i}$,
\begin{widetext}
\begin{eqnarray}
a_{ij}(x)\mathrm{d}x^i \mathrm{d}x^j&=& \frac{g_{rr}\mathrm{d}r^2}{f(r)}+\frac{\beta^2 (r^2+a^2 \cos^2\theta) \mathrm{d}\theta^2}{f(r)}+\frac{1}{f(r)}\left[g_{\varphi\varphi}+\left(\frac{2 M a r\sin^{2}\theta}{r^2+a^2 \cos^{2}\theta}\right)^2\frac{1}{f(r)}\right]\mathrm{d}\varphi^2+\frac{2 g_{r\varphi}}{f(r)}\mathrm{d}\varphi \mathrm{d}r,\\
b_{i}(x)\mathrm{d}x^i &=&-\left(\frac{2 M a r\sin^{2}\theta}{r^2+a^2 \cos^{2}\theta}\right)\frac{\mathrm{d}\varphi}{f(r)},
\end{eqnarray}
in which $V^2=f(r)$ and $f(r)=1-2Mr/(r^2+a^2 \cos^{2}\theta)$. 
In the following, however, we will restrict ourselves to the plane $(r,\varphi)$ by setting $\theta=\pi/2$. Moreover, we note that in our approach, the geodesic of a photon in the plane of the equator remains planar. This leads to the considerably simpler global monopole Randers metric
\begin{equation}\label{16-3}
F\left(r,\varphi,\frac{\mathrm{d}r}{\mathrm{d}t},\frac{\mathrm{d}\varphi}{\mathrm{d}t}\right)=\sqrt{\frac{r^4 \Theta(r,a,\beta) }{(\Delta-a^2)^2}\left(\frac{\mathrm{d}\varphi}{\mathrm{d}t}\right)^2+\frac{r^2 \Xi(r,a,\beta)}{\Delta^2(\Delta-a^2)}\left(\frac{\mathrm{d}r}{\mathrm{d}t}\right)^2+\frac{2 a (1-\beta^2)r^4}{\Delta(\Delta-a^2)}\left(\frac{\mathrm{d}\varphi}{\mathrm{d}t}\right)\left(\frac{\mathrm{d}r}{\mathrm{d}t}\right)}-\frac{2M ar}{\Delta-a^2}\frac{\mathrm{d}\varphi}{\mathrm{d}t},
\end{equation}
where
\begin{eqnarray*}
\Delta(r,a)&=& r^2-2Mr+a^2,\\
\Xi(r,a,\beta)&=&\Delta(r^2-a^2(1-\beta^2))-2Mr(1-\beta^2)a^2,\\
\Theta(r,a,\beta)&=&\beta^2r^2+a^2-2Mr\beta^2 .
\end{eqnarray*}
\end{widetext}
One can easily check that by letting $\beta\to 1$ the Kerr-Randers metric is recovered \cite{werner}.

Now the physical interpretation of this Randers metric $F$ concerning light propagation becomes apparent when we note that, for null curves with $\mathrm{d}s^2=0$, one obtains $\mathrm{d}t=F(x,\mathrm{d}x)$ from the spacetime line element \ref{13-3}. Thus, since Fermat's principle implies that spatial light rays $\gamma$ are selected by stationary arrival time at the observer,
\begin{equation}\notag
0=\delta\,\int\limits_{\gamma}\mathrm{d}t=\delta\,\int\limits_{\gamma_F}F(x, \dot{x})\mathrm{d}t,
\end{equation}
these spatial light rays $\gamma$ are also geodesics $\gamma_F$ of the Randers metric $F$.

The next step is to apply Naz{\i}m's method\cite{nazim} to construct a Riemannian manifold $(\mathcal{M},\bar{g})$ the Randers manifold $ (\mathcal{M}, F) $. This can be done by choosing vector field $\bar{X}$ over $\mathcal{M}$ such that it is smooth and non-zero everywhere (except at single vertex points) and, along the geodesic $\gamma_{F}$, such that $\bar{X}(\gamma_{F})=\dot{x}$. In that case, the Hessian \eqref{10-3} reads
\begin{equation}
\bar{g}_{ij}(x)=g_{ij}(x,\bar{X}(x)),\label{17-3}
\end{equation}

The crucial point to note is that the geodesic $\gamma_{F}$ of $(\mathcal{M}, F)$, is also a geodesic $\gamma_{\bar{g}}$ of $(\mathcal{M},\bar{g})$ i.e. $\gamma_{F}=\gamma_{\bar{g}}$ (see \cite{werner} for details). That is to say, we can use the monopole optical metric to construct the corresponding osculating Riemannian manifold $(\mathcal{M},\bar{g})$ and then compute the deflection angle of light rays in the equatorial plane. Since we are presently interested in the leading terms of the weak deflection limit only, it suffices (cf. \cite{werner}) to take the line $r(\varphi)=b/\sin\varphi $ as approximation of the deflected light ray, where $b$ is the coordinate distance of closest approach to the lens, and use only the leading terms of the vector field $\bar{X}=(\bar{X}^{r}, \bar{X}^{\varphi})(r, \varphi)$ near the light ray, \begin{equation}
\bar{X}^{r}=-\cos\varphi+\mathcal{O}(M,a), \hspace{1cm}\bar{X}^{\varphi}=\frac{\sin^{2}\varphi}{b}+\mathcal{O}(M,a).\label{18-3}
\end{equation}
We shall now see how this construction may be used to compute the  light deflection angle using the Gauss-Bonnet theorem.

\section{Optical curvature and the deflection angle}

The Gauss--Bonnet theorem connects the Riemannian geometry of a surface with its topology. This can be usefully applied to our lensing problem as follows: in the equatorial plane of the osculating optical geometry with metric $\bar{g}$ as defined above, consider a domain $D_R$ bounded by the light ray $\gamma_{\bar{g}}$ and a circular boundary curve $C_R$ centered on the lens and intersecting $\gamma_{\bar{g}}$ in the points $S$ and $O$, which we may regard as some notional light source and observer, both at coordinate distance $R$ from the lens.

Now one can apply the Gauss-Bonnet theorem to this domain $(D_{R},\bar{g})$ with the region $D_ {R} $ with boundary curve $\partial D_{R}=\gamma_{\bar{g}}\cup C_ {R}$ (cf. \citep{werner}),
\begin{equation}\label{19-4}
\iint\limits_{D_{R}}K\,\mathrm{d}S+\oint\limits_{\partial D_{R}}\kappa\,\mathrm{d}t+\sum_{i}\theta_{i}=2\pi\chi(D_{R}),
\end{equation}
where $K$ is the Gaussian curvature and $\kappa=|\nabla_{\dot{\gamma}}\dot{\gamma}|$ is the geodesic curvature, all of course with respect to $\bar{g}$. Furthermore, $\theta_{i}$ are the corresponding exterior angles at the $i$--th vertex. Thus, in our case, we have two exterior jump angles $\theta_{S}$ and $\theta_{O}$ at the vertices $S$ and $O$, respectively. As $R\to \infty$, both jump angles tends to $\pi/2$, hence we have $\theta_{O}+\theta_{S}\to \pi$. Moreover, we note that the Euler characteristic is $\chi(D_{R})=1$ since $D_ {R} $ is non-singular and simply connected. The Gauss--Bonnet theorem can now be recast thus,
\begin{equation}
\iint\limits_{D_{R}}K\,\mathrm{d}S+\oint\limits_{\partial D_{R}}\kappa\,\mathrm{d}t=2\pi\chi(D_{R})-(\theta_{O}+\theta_{S})=\pi.\label{20-4}
\end{equation}

Since the geodesic curvature in the case of geodesics $\gamma_{\bar{g}}$ vanishes i.e. $\kappa(\gamma_{\bar{g}})=0$, we shall now focus on calculating $\kappa(C_{R})\mathrm{d}t$ where $\kappa(C_{R})=|\nabla_{\dot{C}_{R}}\dot{C}_{R}|$. For very large but constant $R$ given by  $C_{R}:= r(\varphi)=R=const$, the radial component of the geodesic curvature reads
\begin{equation}
\left(\nabla_{\dot{C}_{R}}\dot{C}_{R}\right)^{r}=\bar{\Gamma}^{r}_{\varphi \varphi}\left(\dot{C}_{R}^{\varphi}\right)^{2}.\label{21-4}
\end{equation}

Here we note that the first term vanishes, while the second term can be calculated by using the unit speed condition i.e. $\bar{g}_{\varphi \varphi}\,\dot{C}_{R}^{\varphi}\dot{C}_{R}^{\varphi}=1$, and the Christoffel symbol $\bar{\Gamma}^{r}_{\varphi \varphi}$. One can show that for very large but constant radial distance i.e. $r(\varphi)=R=const$, the geodesic curvature reads $\kappa(C_{R})\to R^{-1}$. Meanwhile, for a constant $R$ the monopole optical metric \eqref{16-3} gives
\begin{equation}\label{22-4}
 \mathrm{d}t=\left(\sqrt{\frac{\beta^2 R^2-2MR\beta^2+a^2}{\left(1-\frac{2M}{R}\right)^2}}-\frac{2Ma}{R-2M}\right)\mathrm{d}\,\varphi.
\end{equation}

Now for a very large $R$ the last equation suggests that
\begin{eqnarray}\notag
&&\lim_{R\to\infty}\kappa(C_{R})\mathrm{d}t \\\notag
&=&\lim_{R\to \infty} \left[\sqrt{\frac{\beta^2 -\frac{2M\beta^2}{R}+\frac{a^2}{R^2}}{\left(1-\frac{2M}{R}\right)^2}}-\frac{2Ma}{R^2\left(1-\frac{2M}{R}\right)}\right]\mathrm{d}\,\varphi \\
&=& \beta \,\mathrm{d}\,\varphi.\label{23-4}
\end{eqnarray}

In other words, this result reflects the fact that our original spacetime is globally conical which implies that the optical metric is not asymptotically Euclidean i.e. $\kappa(C_{R})\mathrm{d}t/\mathrm{d}\varphi=\beta\neq 1$. However, this result reduces to the asymptotically Euclidean case, $\kappa(C_{R})\mathrm{d}t/\mathrm{d}\varphi=1$, only if one takes the limit $\beta\to 1$. Now, if we use this result and go back to Eq. \eqref{20-4}, it follows that
\begin{eqnarray}\notag
\iint\limits_{D_{R}}K\,\mathrm{d}S&+&\oint\limits_{C_{R}}\kappa\,\mathrm{d}t\overset{{R\to \infty}}{=}\iint\limits_{
D_{\infty}}K\,\mathrm{d}S \\
&+ & \beta \int\limits_{0}^{\pi+\hat{\alpha}}\mathrm{d}\varphi,\label{17}
\end{eqnarray}
where the domain $D_\infty$ is understood to be an infinite domain bounded by the light ray $\gamma_{\bar{g}}$, excluding the lens, and $\hat{\alpha}$  is the asymptotic deflection angle.

In order to compute the leading orders of the asymptotic deflection angle (cf. \cite{werner}), we can approximate the boundary curve of $D_\infty$ by a notional undeflected ray, that is, the line $r(\varphi)=b/\sin\varphi$, similar to the Born approximation. Then the asymptotic deflection angle from the last equation reduces to
\begin{eqnarray}\notag
\hat{\alpha} &\simeq & \pi\left(\frac{1}{\sqrt{1-8\pi \eta^2}}-1\right)-\frac{1}{\sqrt{1-8\pi \eta^2}}\\
&\times & \int\limits_{0}^{\pi}\int\limits_{\frac{b}{\sin \varphi}}^{\infty}K\,\sqrt{\det \bar{g}}\,\mathrm{d}r\,\mathrm{d}\varphi.\label{25-4}
\end{eqnarray}

Let us now compute the components of $\bar{g}$. Starting from \eqref{16-3} and using equations \eqref{10-3}, \eqref{17-3} and \eqref{18-3}, one finds  that,
\begin{widetext}
\begin{align}\label{19}
\bar{g}_{rr}&=1+\frac{4M}{r}-\frac{2\,M \,a \,r \,\beta^2 \sin^{6}\varphi }{b^3 \left(\cos^{2}\varphi+\frac{r^2 \beta^{2} \sin^{4}\varphi }{b^2}\right)^{3/2}}+\mathcal{O}(M^2,a^2),\\
\bar{g}_{\varphi \varphi}&=\left(r^2+2 M r\right) \beta^2-\frac{2 M a r \left(2 r^2 \beta^2 \sin^4 \varphi +3 b^2 \cos^2 \varphi \right)\beta^2 \sin^2 \varphi }{b^3 \left(\cos^{2}\varphi+\frac{r^2 \beta^{2} \sin^{4}\varphi}{b^2}\right)^{3/2}}+\mathcal{O}(M^2,a^2),\\
\bar{g}_{r\varphi}&=-\frac{\left[(4 M+r)(\beta^2-1)\left(\cos^{2}\varphi+\frac{r^2 \beta^{2} \sin^{4}\varphi }{b^2}\right)^{3/2}-2 M \cos^3 \varphi \right]a}{\left(\cos^{2}\varphi+\frac{r^2 \beta^{2} \sin^{4}\varphi }{b^2}\right)^{3/2} r}+\mathcal{O}(M^2,a^2),
\end{align}
neglecting higher order terms of the angular momentum parameter $a$.  Then the determinant of this metric can be written as
\begin{equation}
\det \bar{g}=\left(r^2+6M r \right)\beta^2-\frac{6  a M  r  \left(r^2 \beta^2 \sin^4 \varphi +b^2 \cos^2 \varphi \right)\beta^2 \sin^2 \varphi }{b^3 \left(\cos^{2}\varphi+\frac{r^2 \sin^{4}\varphi \beta^{2}}{b^2}\right)^{3/2}}+\mathcal{O}(M^2,a^2). \label{22}
\end{equation}

For the Christoffel symbols we find,
\begin{eqnarray}\label{23}
\bar{\Gamma}^{\varphi}_{rr}&=&\frac{2 a M(\beta^2-1)}{\beta^2 r^4}+\frac{aM\cos\varphi \left(6 \beta^2 r^3 \cos^2\varphi \sin^5 \varphi +3 \beta^2 r^3 \sin^7\varphi -8 \beta^2 r^2 b \sin^4 \varphi \cos^2 \varphi -2 b^3 \cos^4\varphi \right)}{b^3 \left(\frac{r^2 \beta^2 \sin^4 \varphi}{b^2}+\cos^2 \varphi \right)^{5/2}r^4 \beta^2 },\\
\bar{\Gamma}^{\varphi}_{r\varphi}&=&\frac{r-M}{r^2}+\frac{aM\sin^2\varphi \left(4 \beta^4 r^4 \sin^8 \varphi +10 \beta^2 b^2 r^2 \sin^4\varphi \cos^2\varphi+3 b^4 \cos^4 \varphi \right)}{b^5 \left(\frac{r^2 \beta^2 \sin^4 \varphi}{b^2}+\cos^2 \varphi \right)^{5/2}r^2}.\label{24}
\end{eqnarray}
\end{widetext}
The Gaussian curvature is
\begin{eqnarray}\label{25}
K&=&\frac{\bar{R}_{r\varphi r\varphi}}{\det \bar{g}}\\\notag
&=&\frac{1}{\sqrt{\det \bar{g}}}\left[\frac{\partial}{\partial \varphi}\left(\frac{\sqrt{\det \bar{g}}}{\bar{g}_{rr}}\,\bar{\Gamma}^{\varphi}_{rr}\right)-\frac{\partial}{\partial r}\left(\frac{\sqrt{\det \bar{g}}}{\bar{g}_{rr}}\,\bar{\Gamma}^{\varphi}_{r\varphi}\right)\right],
\end{eqnarray}
so using the Christoffel symbols and the metric components, we obtain
\begin{equation}
K=-\frac{2M}{r^3}+\frac{3 M a}{r^2}f(r,\varphi,\beta),\label{26}
\end{equation}
with
\begin{widetext}
\begin{eqnarray}\notag
f(r,\varphi,\beta)&=&\frac{\sin^3 \varphi}{b^7 \left(\cos^2\varphi+\frac{r^2 \beta^2 \sin^4\varphi}{b^2}\right)^{7/2}} \Big( 2 \beta^6 r^5 \sin^{11}\varphi+5 \beta^4 b^2 r^3 \cos^2 \varphi \sin^7\varphi-10 \beta^2 b^2 r^3 \cos^4\varphi \sin^5 \varphi \\\notag
&&-9 \beta^2 b^2 r^3 \cos^2\varphi \sin^7 \varphi-\beta^2 r^3 b^2 \sin^9 \varphi+16 \beta^2 b^3 r^2 \cos^4 \varphi \sin^4 \varphi+8 \beta^2 b^3 r^2 \cos^2 \varphi \sin^6 \varphi \\\notag
&&-2 \beta^2 b^4 r \cos^4\varphi \sin^3 \varphi+10 b^4 r \cos^6\varphi \sin \varphi+11 b^4 r \cos^4 \varphi \sin^3 \varphi +4 b^4 r \cos^2 \varphi \sin^5 \varphi \\
&& -4 b^5 \cos^6 \varphi -2 b^5 \cos^4 \varphi \sin^2 \varphi \Big).   \label{34-4}
\end{eqnarray}
\end{widetext}
Now from Eq. \eqref{34-4} and \eqref{25-4}, the asymptotic deflection angle becomes
\begin{eqnarray}
\hat{\alpha}& \simeq & 4\pi^2 \eta^2-\frac{1}{\sqrt{1-8\pi \eta^2}}\\\notag
& \times &\int\limits_{0}^{\pi}\int\limits_{\frac{b}{\sin \varphi}}^{\infty}\left(-\frac{2M}{r^3}+\frac{3 M a}{r^2}f(r,\varphi,\beta)\right)\sqrt{\det \bar{g}}\,\mathrm{d}r\,\mathrm{d}\varphi,
\label{35-4}
\end{eqnarray}
where $b$ is the impact parameter. Evaluating the first integral, we find
\begin{equation}
\int\limits_{0}^{\pi}\int\limits_{\frac{b}{\sin \varphi}}^{\infty} \frac{2M}{r^3}\sqrt{\det \bar{g}}\,\mathrm{d}r\,\mathrm{d}\varphi=\frac{4M}{b}.\label{37-4}
\end{equation}

The second integral can be found by integrating first with respect to the radial coordinate and then make a Taylor expansion around $\eta$, say, up to the fourth order, to find
\begin{eqnarray}\notag
&&\int\limits_{0}^{\pi}\int\limits_{\frac{b}{\sin \varphi}}^{\infty} \frac{3 M a}{r^2}f(r,\varphi,\beta)\sqrt{\det \bar{g}}\,\mathrm{d}r\,\mathrm{d}\varphi \\
&= &\pm\frac{4Ma}{b^2}
\left(1+\frac{12\,\pi \eta^2}{5}
+\mathcal{O}(\eta^4) \right),
\label{38-4}
\end{eqnarray}
in which the positive (resp., negative) sign is for a retrograde (resp., prograde) light ray. Thus, 
\begin{equation}
\hat{\alpha}\simeq 4\pi^2 \eta^2+\frac{4M}{b}\pm \frac{4Ma}{b^2}
+\frac{16 \pi M \eta^2}{b}, \label{39-4}
\end{equation}
as the leading terms, to second order, of the asymptotic deflection angle in the equatorial plane for the rotating global monopole. This result includes, as limiting cases, the standard expression for the Kerr black hole (with $\eta\rightarrow 0$, cf. \cite{werner} and references therein) as well as for the non-rotating global monopole (with $a\rightarrow 0$).
On astrophysical scales, we can also neglect the global monopole mass $M$, which yields the deflection angle of one half of the total deficit angle, i.e. $\hat{\alpha}=\delta/2\simeq 4 \pi^2 \eta^2 $, as expected (cf. \cite{vilenkin}). 
\newline
Finally, our result (\ref{39-4}) obtained by means of the Gauss-Bonnet method can be confirmed with a standard geodesic computation, which also shows the limit of applicability for the approximations used, cf. the appendix.

\section{Rotating Letelier spacetime}

In this section we recall that a Letelier spacetime metric corresponds to a spherically symmetric black hole solution surrounded by a cloud of strings. Firstly, one can write the action for the string cloud model as follows (cf. \cite{Bezerra2}),
\begin{equation}
S=\int_{\Sigma} \mathcal{L}\,\mathrm{d}\xi^{0}\mathrm{d}\xi^{1},
\end{equation}
where $\mathcal{L}$ is the Lagrangian density, while $\xi^{0}$ and $\xi^{1}$ are the timelike and spacelike parameters, and the Lagrangian density can be defined as $\mathcal{L}=M\sqrt{-\frac{1}{2}\Sigma^{\mu\nu}\Sigma_{\mu\nu}}$,
in which $\Sigma^{\mu\nu}$ is a bivector associated with the world sheet of a string and $M$ is a positive string constant. The source of the rotating Letelier spacetime is a cloud of strings described by the following energy-momentum tensor
\begin{equation}
T^{\mu\nu}=\frac{\rho\varSigma^{\mu\lambda}\varSigma_{\lambda}^{\nu}}
{\sqrt{-\frac{1}{2}\Sigma^{\alpha\beta}\Sigma_{\alpha\beta}}},
\end{equation}
where $\rho$ is the proper density. The spherically symmetric solution which describes a non-rotating black hole surrounded by a cloud of strings reads \cite{Bezerra2}

\begin{eqnarray}\notag
\mathrm{d}s^{2}&=&-\left(1-\frac{2M}{r}\right)\mathrm{d}t^{2}+\frac{\mathrm{d}r^{2}}{\left(1-\frac{2M}{r}\right)}+(1-A)r^{2}\\
&\times &\left(\mathrm{d}\theta^{2}+\sin^{2}\theta\mathrm{d}\varphi^{2}\right),\label{49-5}
\end{eqnarray}
with the string cloud parameter $A$. If $A=0$, the above solution corresponds to the Schwarzschild solution. On the other hand, if we let $M=0$, the solution corresponds to a cloud of strings with global topology similar to the global monopole case associated with a solid angle deficit depending on the string cloud parameter $A$. To transform the Letelier spacetime given by Eq. \eqref{49-5} into the corresponding rotating counterpart, one can use the method by Newman and Janis, to obtain \cite{Bezerra1,Bezerra2}
\begin{widetext}
\begin{equation}\label{50-5}
g_{\mu\nu}=\begin{bmatrix}\vspace{0.3cm}-\left(1-\frac{2Mr}{r^{2}+a^{2}\cos^{2}\theta}\right) & 0 & 0 & -\frac{2Mar\sin^{2}\theta}{r^{2}+a^{2}\cos^{2}\theta}\\
\vspace{0.3cm}0 & g_{rr} & 0 & g_{r\varphi}\\
\vspace{0.3cm}0 & 0 & (1-A)(r^{2}+a^{2}\cos^{2}\theta) & 0\\
\vspace{0.3cm}-\frac{2Mar\sin^{2}\theta}{r^{2}+a^{2}\cos^{2}\theta} & g_{\varphi r} & 0 & g_{\varphi\varphi}
\end{bmatrix},
\end{equation}

where
\begin{eqnarray}\notag
g_{rr} & = & \frac{r^{2}-a^{2}\left[A^{2}\sin^{2}\theta-\cos^{2}\theta\right]}{r^{2}-2Mr+a^{2}}-\frac{Aa^{2}\sin\theta\left[2Mr-a^{2}\left(1-\sin^{4}\theta\right)\right]}{\left(r^{2}-2Mr+a^{2}\right)^{2}},\\\notag
g_{r\varphi} & = & \frac{Aa\left[r^{2}\sin^{2}\theta-a^{2}\cos^{2}\theta\left(1+\cos^{2}\theta\right)\right]}{r^{2}-2Mr+a^{2}},\\\notag
g_{\varphi\varphi} & = & \sin^{2}\theta\frac{\left\lbrace (1-A)r^{4}+\left[(1-A)\cos^{2}\theta\right]2a^{2}r^{2}+2Ma^{2}r\sin^{2}\theta+a^{4}\cos^{2}\theta\left((1-A)\cos^{2}\theta+\sin^{2}\theta\right)\right\rbrace }{r^{2}+a^{2}\cos^{2}\theta}.
\end{eqnarray}

Sometimes it is convenient to write the rotating Leteiler spacetime in a more compact form by considering only the terms linear in $a$. By neglecting all terms of order $a^2/r^{2}$ the above metric can be approximated as \cite{Bezerra2}
\begin{equation}
\mathrm{d}s^{2}=-\left(1-\frac{2M}{r}\right)\mathrm{d}t^{2}+\frac{\mathrm{d}r^{2}}{\left(1-\frac{2M}{r}\right)}+(1-A)r^{2}\left(\mathrm{d}\theta^{2}+\sin^{2}\theta\mathrm{d}\varphi^{2}\right)-\frac{4Ma\sin^{2}\theta}{r}\mathrm{d}t\mathrm{d}\varphi+\frac{2\,A\,a\,\sin^{2}\theta}{\left(1-\frac{2M}{r}\right)}\mathrm{d}r\mathrm{d}\varphi.
\label{rotlet}
\end{equation}
\end{widetext}
Notice that we recover the Lense-Thirring spacetime when $A=0$ in the above metric, and that the rotating Letelier spacetime metric can be written in a Schwarzschild-like form but with a solid angle deficit. This enables us to compute light deflection in the rotating Letelier spacetime by performing computations analogous to the method employed in the last section.

\section{Deflection angle by a rotating Letelier spacetime}

Following the same method as in the case of the rotating global monopole,  one can now recast the rotating Letelier metric \eqref{50-5} in the form of \eqref{13-3}. Comparing (\ref{6-2}) and (\ref{rotlet}), we see that the spacetimes correspond when setting $1-A=\beta^2$.  More explicitly, the Randers optical metric for the rotating Letelier spacetime is defined by
\begin{widetext}
\begin{eqnarray}\notag
a_{ij}(x)\mathrm{d}x^i \mathrm{d}x^j&=& \frac{g_{rr}\mathrm{d}r^2}{f(r)}+\frac{(1-A) (r^2+a^2 \cos^2\theta) \mathrm{d}\theta^2}{f(r)}+\frac{1}{f(r)}\left[g_{\varphi\varphi}+\left(\frac{2 M a r\sin^{2}\theta}{r^2+a^2 \cos^{2}\theta}\right)^2\frac{1}{f(r)}\right]\mathrm{d}\varphi^2+\frac{2 g_{r\varphi}}{f(r)}\mathrm{d}\varphi \mathrm{d}r,\\
b_{i}(x)\mathrm{d}x^i &=&-\left(\frac{2 M a r\sin^{2}\theta}{r^2+a^2 \cos^{2}\theta}\right)\frac{\mathrm{d}\varphi}{f(r)},
\end{eqnarray}
where $f(r)=1-2Mr/(r^{2}+a^{2}\cos^{2}\theta)$. Then the  Letelier-Randers optical geometry in the equatorial plane $(r,\varphi)$ for $\theta=\pi/2$ can be represented by the following Finsler metric,
\begin{equation}
F\left(r,\varphi,\frac{\mathrm{d}r}{\mathrm{d}t},\frac{\mathrm{d}\varphi}{\mathrm{d}t}\right)=\sqrt{\frac{r^4 \Theta(r) }{(\Delta-a^2)^2}\left(\frac{\mathrm{d}\varphi}{\mathrm{d}t}\right)^2+\frac{r^2 \Xi(r)}{\Delta^2(\Delta-a^2)}\left(\frac{\mathrm{d}r}{\mathrm{d}t}\right)^2+\frac{2 A a r^4}{\Delta(\Delta-a^2)}\left(\frac{\mathrm{d}\varphi}{\mathrm{d}t}\right)\left(\frac{\mathrm{d}r}{\mathrm{d}t}\right)}-\frac{2M a r}{\Delta-a^2}\frac{\mathrm{d}\varphi}{\mathrm{d}t},
\end{equation}
\end{widetext}
where
\begin{eqnarray}\notag
\Delta(r)&=& r^2-2Mr+a^2,\\\notag
\Xi(r)&=&\Delta(r^2-a^2 A)-2MAa^2r,\\\notag
\Theta(r)&=&(1-A)r^2+a^2-2Mr(1-A).
\end{eqnarray}

Now the metric components of the corresponding osculating Riemannian optical geometry read
\begin{widetext}
\begin{align}\notag
\bar{g}_{rr}&=1+\frac{4M}{r}-\frac{2\,M \,a \,r \,(1-A) \sin^{6}\varphi }{b^3 \left(\cos^{2}\varphi+\frac{r^2 (1-A) \sin^{4}\varphi }{b^2}\right)^{3/2}}+\mathcal{O}(M^2,a^2)\\\notag
\bar{g}_{\varphi \varphi}&=\left(r^2+2 M r\right) (1-A)-\frac{2 M a r \left(2 r^2 (1-A) \sin^4 \varphi +3\, b^2 \cos^2 \varphi \right)(1-A) \sin^2 \varphi }{b^3 \left(\cos^{2}\varphi+\frac{r^2 (1-A) \sin^{4}\varphi}{b^2}\right)^{3/2}}+\mathcal{O}(M^2,a^2)\\\notag
\bar{g}_{r\varphi}&=\frac{\left[A\,(4 M+r)\left(\cos^{2}\varphi+\frac{r^2 (1-A) \sin^{4}\varphi }{b^2}\right)^{3/2}+2 M \cos^3 \varphi \right]a}{\left(\cos^{2}\varphi+\frac{r^2 (1-A) \sin^{4}\varphi }{b^2}\right)^{3/2} r}+\mathcal{O}(M^2,a^2),
\end{align}
\end{widetext}
and the Gaussian curvature becomes
\begin{equation}\notag
K=-\frac{2M}{r^3}+\frac{3 M a}{r^2}f(r,\varphi,A),
\end{equation}
noting that the function $f(r,\varphi,A)$ is  equal to $f(r,\varphi,\beta)$ if we let $\beta=\sqrt{1-A}$. Hence, performing the same computation as above, the asymptotic deflection angle for the rotating Letelier spacetime in the equatorial plane reads, to second order,
\begin{equation}
\hat{\alpha}\simeq \frac{A \pi}{2}+\frac{4M}{b}\pm \frac{4Ma}{b^2}
+\frac{2 M A }{b}.\label{61-6}
\end{equation}

\section{Conclusion}

In this paper, we have calculated the deflection of light by a rotating global monopole black hole and in a rotating Letelier spacetime, in the equatorial plane and in the weak deflection limit, to second order terms. From the Randers optical metrics, we have constructed osculating Riemannian metrics to calculate the metric components, Christoffel symbols, and the Gaussian curvature in both cases. Then we have applied the Gauss-Bonnet theorem to calculate expressions for the asymptotic deflection angle. This is found to increase due to the presence of the rotating global monopole parameter $\eta$ and the strings cloud parameter $A$, respectively. Thus, we have generalized known deflection angles for the Kerr solution and the non-rotating global monopole and Letelier spacetimes. Finally, it would be interesting to see whether one can find these results by using the Gauss-Bonnet theorem intrinsically in Finsler i.e. without resorting to osculating Riemannian geometry.

\section*{Appendix A}
In order to confirm our earlier computation using the Gauss-Bonnet method and exhibit the limits of applicability of the approximations used, we shall study in this appendix the null geodesics using the variational principle $\delta \int \mathcal{L} \,\mathrm{d}s=0$, where $s$ is an affine path parameter. For the sake of simplicity, we will use the stationary metric given by eq. \eqref{6-2}, which is linear in the angular momentum parameter $a$.  The Lagrangian is given by 
\begin{widetext}
\begin{equation}\label{43}
\mathcal{L}=-\frac{1}{2}\left(1-\frac{2M}{r(s)}\right)\dot{t}^2+\frac{\dot{r}^2}{2\left(1-\frac{2M}{r(s)}\right)}+\frac{1}{2}\beta^2 r(s)^2 \left(\dot{\theta}^2+ \sin^{2}\theta \dot{\varphi}^2\right)-\frac{2 M a \sin^{2}\theta}{r(s)} \dot{t} \dot{\varphi}+\frac{a\,(1-\beta^2)\sin^{2}\theta }{\left(1-\frac{2M}{r(s)}\right)}\dot{r} \dot{\varphi}.
\end{equation}
\end{widetext}

As in the main text, we will consider the deflection of light in the equatorial plane $\theta =\pi/2$. Next,  let us define the following two constants of motion $l$ and $\gamma$, given as follows \cite{Boyer}
\begin{eqnarray}\label{44}\notag
p_{\varphi}&=&\frac{\partial \mathcal{L}}{\partial \dot{\varphi}}=\beta^2 r(s)^2 \dot{\varphi}-\frac{2 M a }{r(s)} \dot{t}+\frac{a\,(1-\beta^2) }{\left(1-\frac{2M}{r(s)}\right)}\dot{r}=l   \\
p_{t}&=&\frac{\partial \mathcal{L}}{\partial \dot{t}}=-\left(1-\frac{2M}{r(s)}\right)\dot{t}-\frac{2 M a }{r(s)} \dot{\varphi}=-\gamma.
\end{eqnarray}

Now we can introduce a new variable via $r=1/u(\varphi)$, then it is not difficult to show that the following relation holds
\begin{equation}\label{46}
\frac{\dot{r}}{\dot{\varphi}}=\frac{\mathrm{d}r}{\mathrm{d}\varphi}=-\frac{1}{u^2}\frac{\mathrm{d}u}{\mathrm{d}\varphi}.
\end{equation}

Thus making use of eqs. \eqref{43},\eqref{44} and \eqref{46}, for the null geodesics case, we can express $\dot{\varphi}$ in terms $u$ and the corresponding constants of integration $l$ and $\gamma$. Furthermore, given the freedom to affinely reparametrize null geodesics, one may choose for definiteness $\gamma=1$ \cite{Boyer}. We note that the angle $\varphi$ is measured from the point of closest approach, namely $u(\varphi=0)=u_{max}=1/r_{min}=1/b$ \cite{lorio}, and one can show that for the second constant $l=\beta b$. Then, we get the following differential equation, 
\begin{widetext}
\begin{eqnarray}\notag
\frac{1}{2u^4 (2Mu-1)}\left(\frac{\mathrm{d}u}{\mathrm{d}\varphi}\right)^2 &=& \frac{\left[\Xi(u)+a \frac{\mathrm{d} u}{\mathrm{d} \varphi}\left(\beta^2-1\right)+\beta^2  \right]^2}{2 u^4 \left(2Mau-2Mu\beta b+ \beta b \right) \zeta(u)}+\frac{2Ma \left(\Xi(u)+a \frac{\mathrm{d} u}{\mathrm{d} \varphi}\left(\beta^2-1\right)+\beta^2  \right)}{u \zeta(u)}\\
&+&\frac{\beta^2}{2 u^2}+\frac{a(1-\beta^2)}{u^2 (2Mu-1)}\frac{\mathrm{d}u }{\mathrm{d} \varphi}
\end{eqnarray}
where 
\begin{eqnarray}
\Xi &=& 4M^2 a \beta b u^4 -2 M u^3 a \beta b -2 M \beta u,\\
\zeta &=&  -4 M^2 \beta b u^2 +4 M^2 a u^2+4 M\beta u-2Mu a-\beta b.
\end{eqnarray}
\end{widetext}

We can solve the last equation for $\mathrm{d}u/\mathrm{d}\varphi$, and then to integrate with respect to $u$, to find the deflection angle. By following similar arguments given in Ref. \cite{Boyer}, the deflection angle can be obtained by considering the following integral 
\begin{equation}
\Delta \varphi= 2 \int_0 ^{1/b}  A(u,M,a,\eta,b)  \mathrm{d}u,
\end{equation}
where $A(u,\eta,b)$ is a complicated function which is obtained by considering Taylor expansion series around $\eta$, $M$, and $a$. This function is found to be
\begin{equation}
A(u,\eta,b)=\frac{8 \pi \eta^2 (b^2u^2-1)(2Mu+1)\sqrt{b^2-b^4 u^2}-b\Sigma}{(b^2 u^2-1)\sqrt{(1-b^2 u^2)b^2}},
\end{equation}
where 
\begin{eqnarray}\notag
\Sigma &=&\Big( (4 \pi \eta^2+1)b+Ma u (16 \pi \eta^2 +2)\\
&-& u^2 (4 \pi \eta^2+1)(Mu+1)b^3) \Big).
\end{eqnarray}

Carrying out the integration and keeping only the terms of order $1/b$, and $1/b^{2}$, we can express the final result as 
\begin{equation}
\Delta \varphi =\pi+4 \pi^2 \eta^2+\hat{\alpha},
\end{equation}
where $\hat{\alpha}$ is the deflection angle. As expected, due to the presence of topological defects the term $4 \pi^2 \eta^2$ appears, which of course is not present in the standard Kerr spacetime (cf. \cite{Boyer}). Thus we find the following result for the deflection angle in the weak deflection approximation to second order terms,  
\begin{equation}
\hat{\alpha} \simeq\frac{4M}{b}\pm \frac{4 Ma}{b^2}
+\frac{16 \pi M \eta^2}{b}.
\end{equation}

However, it is important to point out that the term $4 \pi^2 \eta^2$ illustrates the role of global topology in the deflection of light. In this approach, this term is not present in the result for the deflection angle $\hat{\alpha}$. Therefore, in order to find the total deflection angle $\hat{\alpha}_{total}$ we need to add the topological term to $\hat{\alpha}$ i.e. $\hat{\alpha}_{total}=4 \pi^2 \eta^2+\hat{\alpha}$. In the Gauss-Bonnet framework this term naturally appears in the total deflection angle and we don't need to add this term by hand. This result confirms the deflection angle given by Eq. \eqref{39-4}, calculated using the Gauss-Bonnet method under the assumptions stated in the main text.
\newline
Regarding the third order mixed term in $Ma\eta^2$, we may also note that the above geodesic calculation yields $\pm32\pi M a \eta^2/b^2$, whereas the Gauss-Bonnet method with our approximation gives $\pm 128\pi M a \eta^2/5b^2$. This confirms that this approximation is indeed valid to recover the leading order terms in $M$, $\eta^2$, $Ma$ and $M\eta^2$, but would need to be modified, starting from the boundary of the integration domain and the vector field of eq. (\ref{18-3}) used to construct the osculating Riemannian geometry, to correctly reproduce higher order terms such as this one.

\section*{Acknowledgements}

We thank Prof. Gary Gibbons for his helpful comments and constructive suggestions. This work was supported by the Chilean FONDECYT Grant No. 3170035 (A\"{O}).


\begin{thebibliography}{99}

\bibitem{Achucarro} A. Achucarro, B. Hartmann and J. Urrestilla, JHEP {\bf 0507}, 006 (2005).

\bibitem{bao} D. Bao, S. Chern and Z. Shen \textit{An Introduction to Riemann-Finsler Geometry} (Springer, New York, 2000).

\bibitem{Bezerra2} D. Barbosa and V. B. Bezerra, Gen. Rel. Grav. {\bf 48}, 149 (2016).

\bibitem{birrola} M. Barriola and A. Vilenkin, Phys. Rev. Lett.  \textbf{63}, 341 (1989).

\bibitem{Bertrand} B. Bertrand, Y. Brihaye and B. Hartmann, Class. Quantum Grav. {\bf 20} 4495 (2003);
 Y. Brihaye, B. Hartmann and E. Radu, Phys. Rev. D {\bf 74}, 025009 (2006).

\bibitem{Bezerra de Mello} E. R. Bezerra de Mello, Brazilian J. Phys. {\bf 31}, 211 (2001).

\bibitem{Boyer} R. H. Boyer, R. W. Lindquist, J. Math. Phys. \textbf{8}, 265 (1967).


\bibitem{Cheng} H. Cheng and J. Man, Class. and Quantum Grav. {\bf 28}, 015001 (2011).

\bibitem{gibbons3} G. W. Gibbons, Phys. Lett. B \textbf{308}, 237–39 (1993).

\bibitem{gibbons1} G. W. Gibbons and M. C. Werner, Class. Quantum Grav. \textbf{25}, 235009 (2008).

\bibitem{gibbons2} G. W. Gibbons, C. A. R. Herdeiro, C. M. Warnick and M. C. Werner, Phys. Rev. D \textbf{79}, 044022 (2009).

\bibitem{asahi1} A. Ishihara, Y. Suzuki, T. Ono, T. Kitamura and H. Asada, Phys. Rev. D \textbf{94}, 084015 (2016).

\bibitem{asahi2} A. Ishihara, Y. Suzuki, T. Ono and H. Asada, Phys. Rev. D \textbf{95}, 044017 (2017).

\bibitem{Eddington} Joint eclipse meeting of the Royal Society and the Royal Astronomical Society, Observatory \textbf{42}, 389 (1919).

\bibitem{lorio} L. Iorio, Nuovo Cim. B, \textbf{118}, 249 (2003).

\bibitem{kimet} K. Jusufi, Astrophys. Space Sci. \textbf{361}, 24 (2016);
K. Jusufi, Eur. Phys. J. C \textbf{76}, 332 (2016).

\bibitem{Kibble} T. W. B. Kibble, J. Phys. A {\bf 9}, 1387 (1976).

\bibitem{Lee} K. Lee, V. P. Nair and E. Weinberg, Phys. Rev. D {\bf 45}, 2751 (1992);
P. Breitenlohner, P. Forgacs and D. Maison, Nucl. Phys. B {\bf 383}, 357 (1992);
P. Breitenlohner, P. Forgacs and D. Maison, Nucl. Phys. B {\bf 442}, 126 (1995).

\bibitem{liu} L.-H. Liu and T. Prokopec, \url{arXiv:1612.00861}.

\bibitem{Graca} J. P. Morais Gra\c{c}a and V. B. Bezerra, Mod. Phys. Lett. A {\bf 27}, 1250178 (2012).

\bibitem{nazim} A. Naz{\i}m (1936), whose name later changed to A. Naz{\i}m Terzio\u{g}lu (1912-1976). For references, see \cite{werner}.

\bibitem{Newman} E. T. Newman and A. I. Janis, J. Math. Phys. {\bf 6}, 915 (1965).

\bibitem{Schneider} P. Schneider, J. Ehlers and E. E. Falco {\it Gravitational Lenses} (Springer-Verlag, Berlin, 1992); A. O. Petters, H. Levine and J. Wambsganss {\it Singularity Theory and Gravitational Lensing} (Birkh\"{a}user, Boston, 2001).


\bibitem{Bezerra1} R. M. Teixeira Filho and V. B. Bezerra, Phys. Rev D {\bf 64}, 084009 (2001).

\bibitem{vilenkin} A. Vilenkin and E. P. S. Shellard  \textit{Cosmic Strings and Other Topological Defects} (Cambridge University Press, Cambridge, 1994).

\bibitem{Watabe} H. Watabe and T. Torii, Phys. Rev. D {\bf 66}, 085019 (2002).

\bibitem{werner} M. C. Werner, Gen. Rel. Grav. \textbf{44}, 3047 (2012).



\end{thebibliography}
\end{document}